\documentclass[aps,prb,reprint,showkeys,superscriptaddress,floatfix]{revtex4-1}
\usepackage{amsfonts}
\usepackage{amssymb}
\usepackage{amsmath}
\usepackage{hyperref}
\usepackage{graphicx}
\usepackage{bookmark}
\begin{document}
\title{Ultimate Photo-Thermo-Acoustic Efficiency of Graphene Aerogels}
\author{Francesco De Nicola}
\email[E-mail: ]{francesco.denicola@iit.it}
\affiliation{Graphene Labs, Istituto Italiano di Tecnologia, Via Morego 30, 16163 Genova, Italy}
\author{Lorenzo Donato Tenuzzo}
\affiliation{Department of Physics, University of Rome La Sapienza, P.le A. Moro 5, 00185 Rome, Italy}
\author{Ilenia Viola}
\affiliation{CNR NANOTEC-Institute of Nanotechnology, S.Li.M Lab, Department of Physics, University of Rome La Sapienza, P.le A. Moro 5, 00185 Rome, Italy}
\author{Rujing Zhang}
\affiliation{State Key Laboratory of New Ceramics and Fine Processing, School of Materials Science and Engineering, Tsinghua University, Beijing 100084, China}
\author{Hongwei Zhu}
\affiliation{State Key Laboratory of New Ceramics and Fine Processing, School of Materials Science and Engineering, Tsinghua University, Beijing 100084, China}
\author{Augusto Marcelli}
\affiliation{INFN-Laboratori Nazionali di Frascati, Via Enrico Fermi 40, 00044 Frascati (RM), Italy}
\affiliation{RICMASS, Rome International Center for Materials Science Superstripes, Via dei Sabelli 119A, 00185 Rome, Italy}
\author{Stefano Lupi}
\affiliation{Graphene Labs, Istituto Italiano di Tecnologia, Via Morego 30, 16163 Genova, Italy}
\affiliation{Department of Physics, University of Rome La Sapienza, P.le A. Moro 5, 00185 Rome, Italy}
\date{\today}
\maketitle
\indent \textbf{The ability to generate, amplify, mix, and modulate sound with no harmonic distortion in a passive opto-acoustic device would revolutionize the field of acoustics. The photo-thermo-acoustic (PTA) effect allows to transduce light into sound without any bulk electro-mechanically moving parts and electrical connections, as for conventional loudspeakers. Also, PTA devices can be integrated with standard silicon complementary metal-oxide semiconductor (CMOS) fabrication techniques. Here, we demonstrate that the ultimate PTA efficiency of graphene aerogels, depending on their particular thermal and optical properties, can be experimentally achieved by reducing their mass density. Furthermore, we illustrate that the aerogels behave as an omnidirectional point-source throughout the audible range with no harmonic distortion. This research represents a breakthrough for audio-visual consumer technologies and it could pave the way to novel opto-acoustic sensing devices.}\\
\begin{figure}[hb!]
\centering
\includegraphics[width=0.45\textwidth]{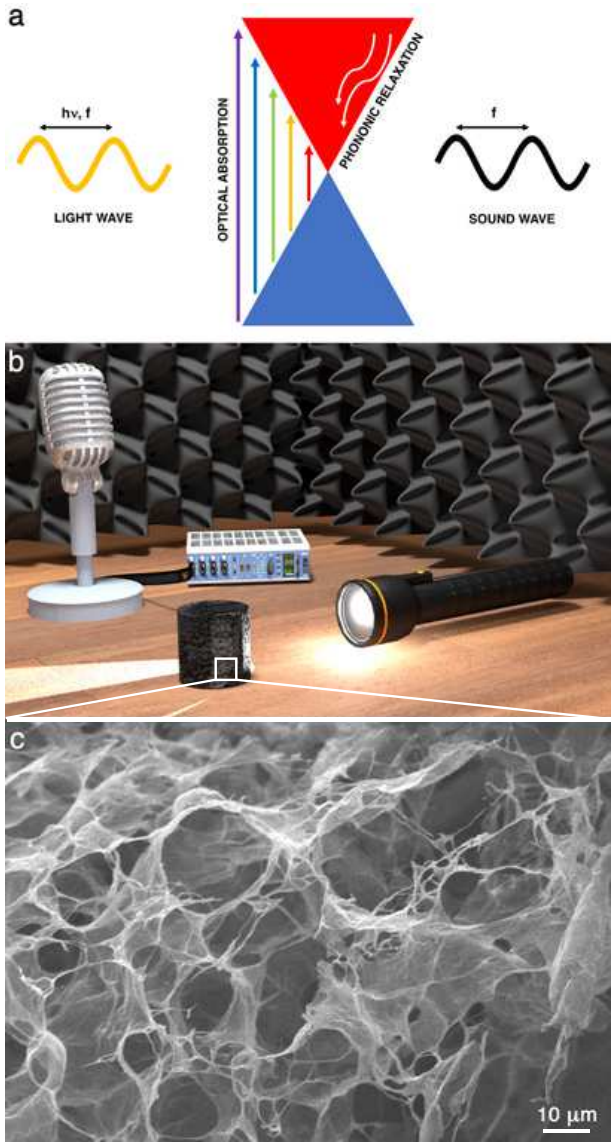}
\caption{Photo-thermo-acoustic effect in graphene aerogels. \textbf{a}, Scheme of the PTA microscopic mechanism in graphene. Light with photon energy $h\nu$ modulated in intensity at a frequency $f$ is absorbed by graphene. Excited electrons relax by ultrafast electron-phonon scattering processes, which heat the sample. The temperature oscillation leads to a pressure wave with the same frequency $f$, i.e. sound. \textbf{b}, Conceptual scheme of the experiment. A panchromatic light is shone on the sample. The sound emitted by the aerogel is recorded by a microphone connected to a sound card. \textbf{c}, Representative SEM micrograph of a graphene aerogel.}
\label{fig:Figure1}
\end{figure}
\begin{figure*}[ht]
\centering
\includegraphics[width=1\textwidth]{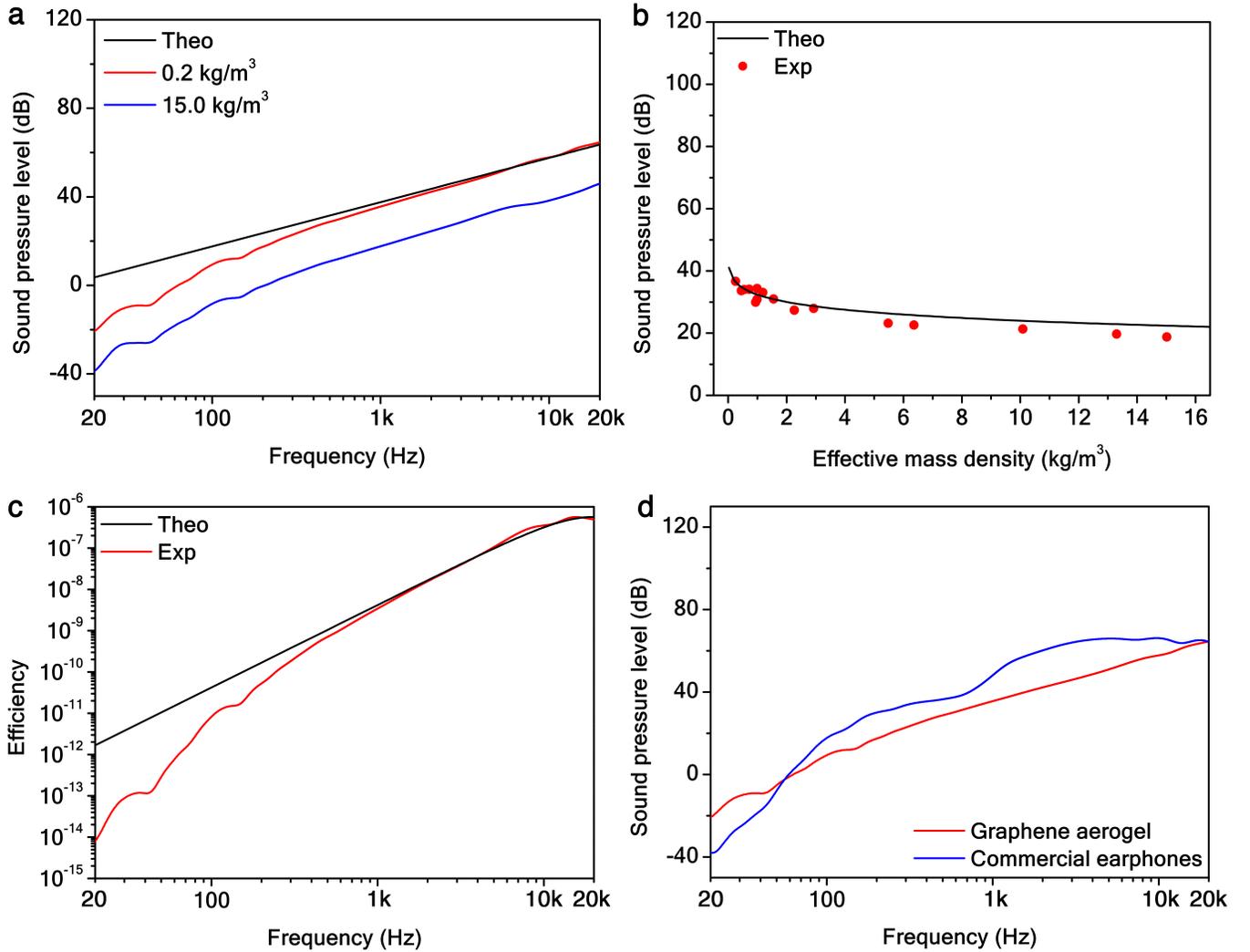}
\caption{Density-dependent PTA frequency response in graphene aerogels. \textbf{a}, Unweighted SPL frequency response at the input power of 1 W and recorded at 1 m distance from the source for two graphene aerogels with effective mass density $\rho_{s}=$0.25 kg/m$^{3}$ (red solid curve) and $\rho_{s}=15.00$ kg/m$^{3}$ (blue solid curve). The black solid curve represents the limiting analytical PTA model with no free parameters. \textbf{b}, Unweighted SPL at 1 W/1 m/1 kHz as a function of the aerogel mass density (red dots). The black solid curve represents the PTA model. \textbf{c}, Unweighted PTA efficiency for the aerogel with effective mass density $\rho_{s}=0.25$ kg/m$^{3}$ (red solid curve). The black solid curve represents the PTA model. \textbf{d}, Unweighted SPL frequency response at the input power of 1 W and recorded at 1 m distance from the source for a graphene aerogel with effective mass density $\rho_{s}=$0.25 kg/m$^{3}$ (red solid curve) and commercial earphones (blue solid curve).}
\label{fig:Figure2}
\end{figure*}
\begin{figure}[ht]
\centering
\includegraphics[width=0.5\textwidth]{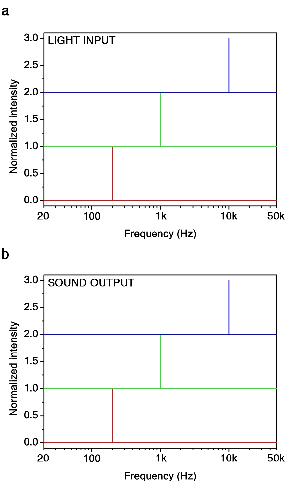}
\caption{Harmonic analysis of PTA sound emission in graphene aerogels. Fast Fourier transform of the light input (\textbf{a}) and sound output (\textbf{b}) signals at 200 Hz, 1 kHz, and 10 kHz for the aerogel with effective mass density $\rho_{s}=0.25$ kg/m$^{3}$. The input and output signals are undistorted in the whole audible range.}
\label{fig:Figure3}
\end{figure}
\section*{Introduction}
\label{sec:intro}
\indent The photo-thermo-acoustic (PTA) effect \cite{Tian2015,Giorgianni2018,Rosencwaig1976} is a two-stage process: a photo-thermal effect followed by a thermo-acoustic effect \cite{Vesterinen2010,Hu2010,Arnold1917}. From a microscopic perspective, when light with photon energy $h\nu$ modulated in intensity at a frequency $f$ shines on a material, the absorbed photons excite the electrons by intraband or interband transitions. Such excited states relax to the ground state by emitting optical and acoustic phonons. It has been well-established that the energy relaxation of energetic charge carriers is mediated by electron-phonon interactions, leading to a hot lattice temperature and then to an efficient photo-thermal process occurring on the order of picoseconds \cite{Ashcroft1976,Bonaccorso2010}. Then, the material adiabatically heats the ambient air molecules at the frequency $f$. Such a temperature oscillation produces a density oscillation in the gas by thermal contraction and expansion of a thin layer of air, which accordingly varies its pressure above and below the ambient atmospheric pressure, leading to a coherent generation of longitudinal sound waves. This second stage is the thermo-acoustic process. In Figure \ref{fig:Figure1}a, the PTA effect is sketched for graphene, for example.\\
\indent The root-mean-square sound pressure amplitude $p_{rms}$ of a PTA wave can be derived by a general analytical solution of the PTA model (see Supplementary Information 1), as follows
\begin{equation}
p_{rms}=\frac{fQ_{0}(\lambda)}{\sqrt{2}r_{0}C_{p,g}T_{g}}\frac{e_{g}}{M(f)e_{s}+e_{g}}\mathcal{D}(\theta,\phi),
\end{equation}
with $f$ the frequency of sound, $Q_{0}(\lambda)$ the amplitude of the power of the light with wavelength $\lambda$ absorbed by the material, $r_{0}$ the far-field distance from the sound source, $e_{i}=\sqrt{k_{i}\rho_{i}C_{p,i}}$ the thermal effusivity of the material $(i=s)$ and the medium $(i=g)$, with $k_{i}$ the thermal conductivity, $\rho_{i}$ the mass density, and $C_{p,i}$ the specific heat capacity at constant pressure, $T_{g}$ the medium (air) temperature, $M(f)\approx1$ a frequency-dependent factor, and $\mathcal{D}(\theta,\phi)$ the far-field directivity. From the above equation is evident that efficient thermo-acoustic sound generation can be achieved by meeting a single condition: heat should flow efficiently to the surrounding medium (i.e., the material effusivity $e_{s}$ should match air effusivity $e_{g}$). The condition can be satisfied by employing a very thin material with low heat capacity per unit area (HCPUA) $\zeta_{s}=L_{s}\rho_{s}C_{p,s}$, where $L_{s}$ is the material thickness, therefore high thermal conductivity $k_{s}$, for instance graphene \cite{Tian2011,Tian2012,Suk2012,Xu2013,Tian2014,Tian2015,Kim2016a,Tao2017,Heath2017}; or a porous material with high HCPUA and low $k_{s}$, for example aerogels \cite{Giorgianni2018,Kim2016,Fei2015,Aliev2015}. However, in the former case heat conduction to air is typically hindered by heat loss in the material supporting substrate. Furthermore, ultrablack \cite{DeNicola2017} materials with quasi-blackbody absorptivity $a(\lambda)=0.99$ represent the best choice for PTA sound generation. This should rule out the use of two-dimensional materials, due to their low absorptivity. For instance, single-layer graphene has a constant low absorptivity $a(\lambda)=0.03$ from the infrared to the visible range \cite{Bonaccorso2010}. Moreover, conductors over semiconductors and insulators are preferred, owing to the their gapless optical spectrum.\\
\indent Here, we demonstrate that a graphene aerogel having $a(\lambda)=0.99$ can emit omnidirectional sound with no harmonic distortion by PTA effect. In particular, we experimentally achieved the ultimate PTA efficiency of such aerogels by tuning their mass density.\\
\section*{Results and Discussion}
\label{sec:results}
\indent The scheme of the experimental setup is reported in Figure \ref{fig:Figure1}b (see Methods). Graphene aerogels were synthesized by a solvothermal process of reduced graphene oxide \cite{Wu2015}. The samples have a transversal size $L_{x}=1.5-2$ cm and a thickness $L_{s}=0.2-1$ cm, with a mass $m=0.2-15$ mg. From the scanning electron microscopy (SEM) micrographs in Figure \ref{fig:Figure1}c, it is possible to observe the fractal random network of graphene flakes that constitute the aerogels. Such a porous system cannot be treated as an homogeneous solid, therefore the effective medium approximation must be considered in order to estimate its mass density. By contact angle measurements (see Supplementary Information 2), we carefully estimated an aerogel porous fraction $\Phi_{air}=0.43-0.92$, leading to an effective mass density $\rho_{s}$ down to $0.249\pm0.01$ kg/m$^{3}$ (see Supplementary Information 2). Furthermore, all the samples exhibit a reflectivity $R(\lambda)<0.01$ from the infrared to the ultraviolet range, hence an absorptivity $a(\lambda)\equiv1-R(\lambda)=0.99$ (see Supplementary Information 3). The aerogel average heat capacity is $C_{s}=10^{-2}\pm10^{-3}$ J/K, while their average thermal conductivity and diffusivity are $k_{s}=10^{-2}\pm10^{-3}$ W/mK and $\alpha_{s}=10^{-6}\pm10^{-7}$ m$^{2}$/s, respectively (see Supplementary Information 4). Therefore, the minimum effusivity obtained for the samples is $e_{s}=7.27\pm0.03$ Ws$^{1/2}$/m$^{2}$K, which matches well the value for air $e_{g}=6$ Ws$^{1/2}$/m$^{2}$K \cite{Callen1985}.\\
\indent In Figure \ref{fig:Figure2}a, the sound frequency response at the input power of 1 W and recorded at 1 m distance from the source, according to the Audio Engineering Society standard for acoustic measurements (AES02-1984-r2003), for two graphene aerogels with different mass density is shown. It can be observed that the ultimate theoretical sound pressure level \cite{Everest2001} (SPL) $L_{p}=20\log_{10}{(p_{rms}/p_{0})}$, with $p_{0}=20$ $\mu$Pa the root-mean-square pressure of the minimum audible threshold, is achieved for the lower density. Experimental data departs from the PTA analytical model (see Supplementary Information 1) with no free parameters below 200 Hz, where the modal resonances due to the finite size of the sound-proof room decrease the signal-to-noise ratio. Moreover, by fitting the curve with the PTA model, leaving the sample effusivity as the only free parameter, the fit returns $e_{s}=7.47\pm0.02$ Ws$^{1/2}$/m$^{2}$K, which is in great agreement with the value obtained by the thermal characterization  of the samples. In Figure \ref{fig:Figure2}b, the SPL dependence on the sample density at 1 W/1 m/1 kHz is illustrated along with the PTA model with no free parameters. The relation between SPL and density sets an asymptotic density $\rho_{0}\equiv e_{g}^{2}/k_{s}C_{p,s}=0.2$ kg/m$^{3}$ for which the material effusivity matches that of the medium.\\
\indent The PTA efficiency (i.e., the ratio between the output acoustic power and input light power) at 1 W/1 m for the graphene aerogel with lower density along the limiting theoretical curve (see Supplementary Information 1) are reported in Figure \ref{fig:Figure2}c. For instance, we achieved an ultimate efficiency $\eta=10^{-8}$ at 1 W/1 m/1 kHz. On the other hand, the thermodynamic efficiency of the PTA process $\eta^{\prime}\equiv1-T_{c}/T_{h}=40\%$ is given by a Carnot cycle \cite{Vesterinen2010,Aliev2015} between two heat reservoirs of temperature $T_{c}=300$ K and $T_{h}=500$ K. Since graphene aerogels behave as a quasi-blackbody, evidently 61\% of the radiative heat provided is emitted in the environment without generation of sound. However, the relevant figure of merit for loudspeakers is not the efficiency but the sensitivity \cite{Everest2001} (i.e., the SPL relative to the maximum audible threshold of 120 dB). Despite the PTA efficiency is rather low for the aerogels, their sensitivity is -80 dB that is well above the minimum audible threshold (-120 dB) \cite{Everest2001} and it is the highest reported for thermo-acoustic devices (Supplementary Information 5). By comparison, commercial earphones have an electro-acoustic efficiency about $10^{-7}$ and a sensitivity of -72 dB at 1 W/1 m/1 kHz (Figure \ref{fig:Figure2}d), which is close to the value for graphene aerogels due to their similar surface area. Therefore, graphene aerogels could be successfully used as PTA earphones. The sensitivity of graphene aerogels could be further increased in order to exploit them as loudspeakers, for instance by increasing their surface area so that the sound emitted by every point on the illuminated surface coherently adds up to increase the SPL.\\
\begin{figure}[ht]
\centering
\includegraphics[width=0.5\textwidth]{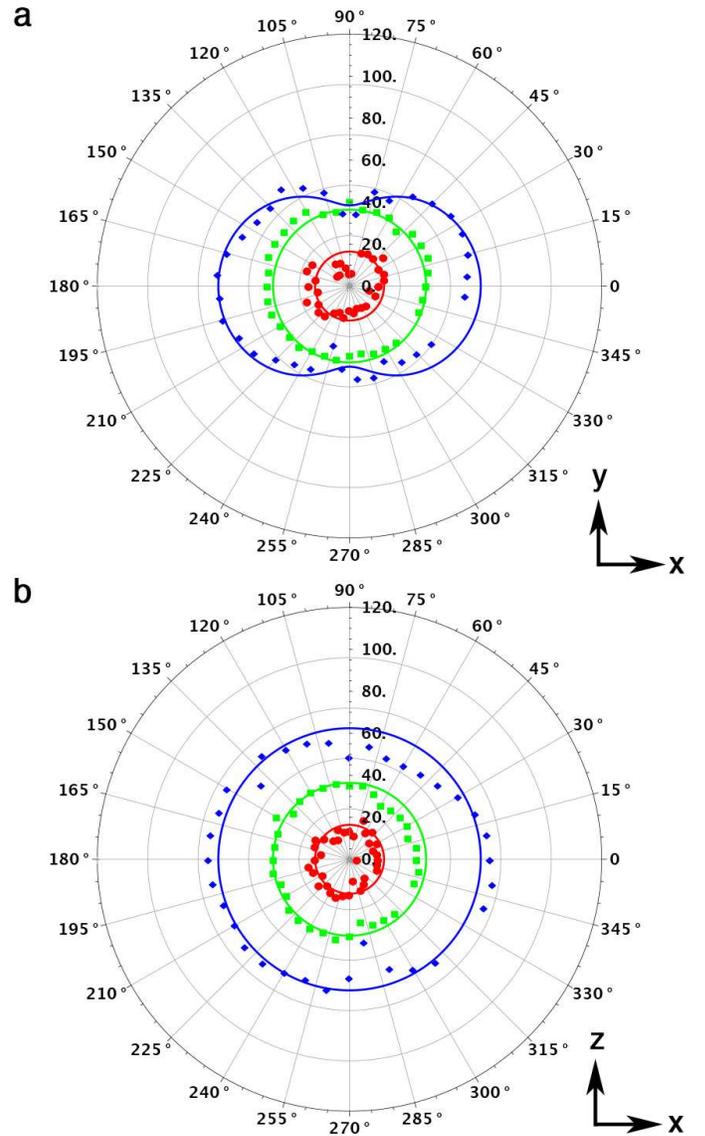}
\caption{Photo-thermo-acoustic directivity in graphene aerogel loudspeakers. Experimental directivity of the sound emitted at 1 W/1 m and at 200 Hz (red dots), 1 kHz (green squares), and 10 kHz (blue diamonds) from the graphene aerogel with effective mass density $\rho_{s}=$0.25 kg/m$^{3}$ in the azimuthal (\textbf{a}) and polar (\textbf{b}) planes. Solid curves represents the PTA model. The directivity indexes \cite{Beranek1993} are DI(200 Hz)=11 dB, DI(1 kHz)=13 dB, and DI(10 kHz)=18 dB.}
\label{fig:Figure4}
\end{figure}
\indent In Figure \ref{fig:Figure3} (a, b), the frequency response of the input (light) and output (sound) signals are respectively depicted in the bass (200 Hz), mid (1 kHz), and treble (10 kHz) range in order to estimate the total harmonic distortion \cite{Everest2001} (THD) of the sound generated from the graphene aerogel with the lowest density. We observed no THD in graphene aerogels over the whole audible range. In particular, no harmonic distortion is present in the treble range, where the human voice frequencies are typically louder \cite{Everest2001}. It is worth noting that the PTA effect does not provide any second harmonic generation, likewise the thermo-acoustic effect does due to Joule heating \cite{Heath2017}. This is because the sample heating is radiative (see Supplementary Information 4). By comparison, commercial hi-fi loudspeakers and earphones typically have a THD about 1\% (see Supplementary Information 5).\\
\indent In Figure \ref{fig:Figure4} (a, b), the theoretical and experimental directivity \cite{Beranek1993} of the sound emitted at 1 W/1 m from the graphene aerogel with lowest density is reported for the bass (200 Hz), mid (1 kHz), and treble (10 kHz) ranges in the azimuthal and polar planes, respectively. The samples behave as an omnidirectional point source up to 20 kHz, beyond which they present a dipolar pattern. Since there are no electro-mechanically moving parts, the sound is equally generated with no destructive interference from both the sides of the illuminated spot of the aerogel, which acts as a diaphragm of thermal thickness $\mu=60$ $\mu$m-1 mm (see Supplementary Information 1), depending on the sound frequency. Therefore, no bass reflex technique \cite{Beranek1993} is needed in PTA loudspeakers.\\
\indent In conclusion, graphene aerogels can be employed as PTA loudspeakers with omnidirectional emission and no harmonic distortion in the audible range. We demonstrated that the theoretical limiting efficiency of such loudspeakers can be achieved simply by tailoring their mass density. Therefore, graphene aerogel loudspeakers may offer a valid alternative to commercial electro-mechanical earphones and in general hi-fi loudspeakers, provided that the illuminated spot is sufficiently large. Moreover, we believe that PTA loudspeakers could pave the way to novel opto-acoustic sensing devices and metrology applications.
\section*{Methods}
\label{sec:methods}
The experimental setup was placed in a sound-proof room ($3\times2\times2$ m) in order to acoustically insulate the experimental apparatus from environmental noise and to reduce internal sound reflections. The audio signal flow was calibrated in an anechoic chamber at the National Institute for Insurance against Accidents at Work (INAIL). Light was generated by a panchromatic LED source (Thorlabs) controlled by a custom LED driver connected to a PC by RME Fireface UFX sound card. Light was modulated in intensity by a sine wave generated by Room EQ Wizard (REW) software. The emitted light was focused on the graphene aerogel and the generated sound was acquired by a calibrated microphone (Earthworks M50) connected to the sound card and placed in front of the sample at a fixed distance of $r_{0}=1$ cm for near-field measurements and $r_{0}=1$ m for far-field measurements. Gated measurements \cite{Everest2001} were carried out in order to record the sample sound frequency response without any sound reflections, as in anechoic chamber. The sample dynamic range acquired as a function of the sound frequency, was divided by the background noise of the room and smoothed at 1/3 per octave. Therefore, here the SPL corresponds to the signal-to-noise ratio. Directivity patterns were measured by a goniometric stage with the sample fixed in the center, the microphone rotating around the sample, and the LED placed outside the stage. The commercial earphones (Apple earpods) and the commercial (Soundvision) loudspeakers were driven by a sine wave generated by REW software.

\begin{acknowledgments}
The authors acknowledge that this project has received funding from the European Union's Horizon 2020 research and innovation programme under grant agreement No. 785219 - GrapheneCore2. The authors also acknowledge that this project has received the financial support of the Bilateral Cooperation Agreement between Italy and China of the Italian Ministry of Foreign Affairs and of the International Cooperation (MAECI) and the National Natural Science Foundation of China (NSFC), in the framework of the project of major relevance 3-Dimensional Graphene: Applications in Catalysis, Photoacoustics and Plasmonics. The authors also acknowledge D. Annesi for the audio signal flow calibration in anechoic chamber at the National Institute for Insurance against Accidents at Work (INAIL).
\end{acknowledgments}
\section*{Author contribution}
\label{sec:AuthorContribution}
\noindent F.D.N., L.D.T., I.V., R.J.Z., H.W.Z., A.M., and S.L. conceived the experiments. F.D.N. and L.D.T. performed the acoustic, thermal, electrical, and density characterization of the samples and data analysis. F.D.N. and I.V. carried out the contact angle measurements. F.D.N. computed the photo-thermo-acoustic simulations. R.J.Z. and H.W.Z. prepared the graphene samples. F.D.N. and S. L. wrote the paper with the contribution of all the authors.\\\\
The authors declare no competing interests.
\end{document}